# Polaron Coherence as Origin of the Pseudogap Phase in High Temperature Superconducting Cuprates

A. Bussmann-Holder[1], H. Keller[2], A. R. Bishop[3], A. Simon[1], and K. A. Müller[2]

[1]Max-Planck-Institut für Festkörperforschung, Heisenbergstr. 1, D-70569 Stuttgart, Germany

[2]Physik Institut der Universität, Winterthurerstr. 190, CH-8057 Zürich, Switzerland

[3]Los Alamos National Laboratory, Theoretical Division, Los Alamos, NM87545, USA

Within a two-component approach to high $T_c$ copper oxides including polaronic couplings, we identify the pseudogap phase as the onset of polaron ordering. *This ordering persists in the superconducting phase.* A huge isotope effect on the pseudogap onset temperature T* is predicted and in agreement with experimental data. The anomalous temperature dependence of the mean square copper – oxygen ion displacement observed above, at and below $T_c$ stems from an s-wave superconducting component of the order parameter, whereas a pure d-wave order parameter alone can be excluded.

The phase diagram of high temperature superconducting (HTS) copper oxides includes antiferromagnetism at zero or small hole doping followed by a spin glass phase, which coexists for a limited doping regime with superconductivity, and finally superconductivity alone at larger doping. Overlaid is the so called pseudogap phase which, for much of the doping regime, onsets at much higher temperatures T* than the other phases. The pseudogap phase is rather ill-defined since it is identified in various ways in different experiments, depending on the properties probed by specific experiments, and by the spatial and temporal resolution of the probes. Theoretically, till now, no consensus about the origin of this phase has been reached, and, in addition, it remained unclear which role this phase plays for HTS's. Here we show that polaron formation is the origin of this phase where their coherence defines its onset temperature T*. This phase is anticipated through the slowing down of the combined electron – lattice displacement which starts far above T*.

The involvement of electron – lattice interactions in the pseudogap phase formation has been firmly established experimentally by two different techniques and on two different cuprate families by showing that a large isotope effect on T* is present [1, 2] which is sign-reversed as compared to the one on the superconducting transition temperature $T_c$. These experiments, which probe charge ordering are, however, in contrast to experiments which probe the spin channel, as, e.g., NQR [3], where the isotope effect on T* is small and has the same sign as the one on $T_c$. Both data together thus provide compelling evidence that the pseudogap phase is a multi-component phase where spin and charge ordering coexist. Another important indication of strong and unconventional electron-lattice interactions stems from EXAFS and ion channeling data [4 – 7], where strong local deviations in the Cu-O bond lengths from the average one have been observed. The Cu-O-Cu bond length splits into two different lengths separated by ≈0.06Å. This splitting is probably dynamical and can be associated with a local double-well potential reminiscent of ferroelectricity [8]. These anomalies appear first at T* and are characterized by an unusual upturn in the mean square relative copper – oxygen displacement $\sigma^2(T)$. A similar but less pronounced peak of $\sigma^2(T)$ appears at $T_c$ followed by a rapid decrease in the superconducting phase.

That the superconducting phase is connected to polaron, respectively bipolaron [9], formation has been suggested in a variety of early [10, 11] and recent work [12]. In particular, the isotope effect on $T_c$, the magnetic penetration depth λ [12], and the superconducting energy gap [13] have been consistently explained in terms of polaronic band renormalizations [14]. Interestingly, this approach has been studied almost exclusively in terms of consequences arising for the electronic structure. Its relevance to the lattice degrees of freedom has generally not been the focus since polaron formation leads initially to a rigid frequency shift of the lattice harmonic oscillators. Only studies on very small systems have been performed early on, concentrating on nonadiabaticity, anharmonicity and the special role played by the apical oxygen ion [15 – 17].

The recent EXAFS data have motivated us to explore in more detail here how the lattice is affected by its coupling to the charge degrees of freedom. The starting point of this study is the electron-lattice interaction Hamiltonian defined by [12]:

$$H = \sum_i \varepsilon(i) c_i^+ c_i + \sum_{q,j} \hbar \omega_{q,j}^{(0)} (b_{q,j}^+ b_{q,j} + \frac{1}{2}) - \sum_i c_i^+ c_i \sum_{q,j} \hbar \omega_{q,j}^{(0)} \gamma_i(q)(b_q^+ + b_q) - \sum_{i,m} c_m^+ c_i \sum_{q,j} \tilde{\gamma}_{i,m}(q)(b_{q,j} + b_{q,j}^+). \quad (1)$$

Here $c^+, c$ are site $i$ dependent electron creation and annihilation operators and with energies $\varepsilon$, and $b^+, b$ are phonon creation and annihilation operators with momentum $q$ and branch $j$ dependent bare phonon energy $\omega_{q,j}^{(0)}$. The coupling between electron and lattice degrees of freedom is given by $\gamma$ and $\tilde{\gamma}$, respectively, where the former is the diagonal coupling and the latter the off-diagonal coupling, which we take as a fraction of the diagonal one. In order to analyze this Hamiltonian, a decoupling scheme for the electronic and ionic degrees of freedom in terms of the Lang-Firsov [18] canonical transformation is applied, corresponding to:
$\tilde{H} = e^{-S} H e^S$, where $S = \sum_{i,q} c_i^+ c_i (b_q^+ - b_q) \gamma_i(q)$. With this choice, exact relations for the transformed creation and annihilation operators are obtained:

$$\tilde{c}_i = c_i \exp[\sum_q \gamma_i(q)(b_q^+ - b_q)]; \quad \tilde{c}_i^+ = c_i^+ \exp[-\sum_q \gamma_i(q)(b_q^+ - b_q)] \quad (2a)$$

$$\tilde{b}_q = b_q + \sum_q \gamma_i(q) c_i^+ c_i; \quad \tilde{b}_q^+ = b_q^+ + \sum_q \gamma_i(q) c_i^+ c_i. \quad (2b)$$

From Eq. (2a), important renormalizations in the charge degrees of freedom are obtained, since the exponential band narrowing carries an isotope effect [12, 19, 20]. As has been shown recently, these renormalizations explain the doping dependent isotope effect on $T_c$ and the magnetic penetration depth, as well as the scaling between isotope effects on the superconducting energy gaps and their related $T_c$'s [13]. In addition to these important consequences for superconductivity, a level shift in the band energies appears, which in a multiband model leads to energy level adjustments that facilitate interband interactions [19].

Here we concentrate on the correlated effect arising from the lattice degrees of freedom and evaluate its response to the polaron formation. This means that the lattice related Hamiltonian transforms to:

$$\tilde{H} = \sum_{q,j} \hbar \tilde{\omega}_{q,j} (\tilde{b}_{q,j}^+ \tilde{b}_{q,j} + \frac{1}{2}), \qquad (3)$$

where the momentum $q$ and branch $j$ dependent renormalized frequencies $\tilde{\omega}_{q,j}$ are given by:

$$\tilde{\omega}_{q,j}^2 = \omega_{q,j}^{(0)2} - \gamma_j^2(q) \sum_k \frac{1}{\varepsilon(k)} \tanh \frac{\varepsilon(k)}{k_B T}, \qquad (4)$$

with $\varepsilon(k)$ being the Fourier transform of the site representation and $\omega_{q,j}^{(0)}$ the bare unrenormalized frequency. The coupling to the electronic degrees of freedom introduces an important temperature dependent softening of this coupled mode which is now no longer a pure lattice mode but represents the combined distortion of lattice and electronic degrees of freedom. The momentum $k$ dependent electronic dispersion is given by the LDA [21] derived form:

$$\varepsilon(k) = -2t_1(\cos k_x a + \cos k_y b) + 4t_2 \cos k_x a \cos k_y b + 2t_3(\cos 2k_x a + \cos 2k_y b)$$
$$\mp t_4(\cos k_x a - \cos k_y b)^2 / 4 - \mu, \qquad (5)$$

where $t_1$, $t_2$, $t_3$ account for nearest, second nearest and third nearest neighbor hopping integrals in the $CuO_2$ planes. $t_4$ is the interlayer hopping between the $CuO_2$ planes, and $\mu$ is the chemical potential which controls the band filling. Below $T_c$, $\varepsilon(k)$ in Eq. (4) has to be replaced by $E(k) = \sqrt{\varepsilon(k)^2 + \Delta(k)^2}$, where $\Delta(k)$ is the superconducting energy gap which can be of d-wave or s-wave symmetry, or be represented by a mixed s+d wave order parameter as suggested from recent μSR experiments [22 – 24]. As mentioned above, $\tilde{\omega}_{q,j}^2$ gains a substantial temperature dependence due to its coupling to the charge and softens with decreasing temperature at finite momentum which defines the periodicity of a modulated structure.

When this softening of $\tilde{\omega}_{q,j}^2$ is complete, a dynamic superstructure in the polaron spatial distribution appears, which we identify here with the so-called stripe pseudo-gap phase [4]. Within this description, the onset temperature is determined by the coupling constant $\gamma$ and the energy of the unrenormalized mode frequency $\omega_{q,j}^{(0)}$ which is given by the implicit relation:

$$\frac{\omega_{q,j}^{(0)2}}{\gamma_j^2(q)} = \int \frac{dk}{\varepsilon(k)} \tanh \frac{\varepsilon(k)}{k_B T^*} \qquad (5)$$

Here we assume that $\gamma$ is the relevant control parameter which systematically grows upon approaching the underdoped regime, where localization sets in and a metal to insulator transition takes place. Note that T* is isotope dependent through the coupled effects of the isotope dependence of the bare lattice frequency $\omega_{q,j}^{(0)}$ and the polaronic renormalizations of the band energies. The isotope effect on T* is sign reversed as compared to the one on $T_c$, and it is huge. This finding is in full agreement with neutron scattering data [1] and EXAFS results [2] and demonstrates the importance of polaron formation.

Above T* the polarons are transient, dynamic and randomly distributed over the lattice forming around the doped hole, since the extra charge introduced by doping induces a *local* lattice distortion which is tied to this charge. At high temperatures the dynamics of these objects exhibit rather high frequencies. Upon approaching T* their dynamics slow down and are almost frozen at T* according to Eqs. 4, 5. Below T* the polarons become persistent but are still dynamic with high frequencies (see figure 1) and confined to the new patterned modulations which are determined by the momentum at which the local dynamics freeze in. In this phase the polaron mode differs from the one above T*, since charge and lattice displacement now form a collective coherent mode (Eqs. 2a, 2b) [4]. The temperature dependence of this polaronic mode frequency is shown in Fig. 1 for $\gamma^2 = 0.25$. Other values of $\gamma$ do not change the overall T-dependence. Above T*, typical mode softening is observed (Eq. 4). It has to be kept in mind, however, that this mode is again not a simple lattice mode, but is determined by the electron and lattice displacement coupling (Eqs. 2a, 2b). *Also, it is important to note that this new mode is local, restricted to a few lattice sites*. Since this mode originates from the bare lattice mode $\omega_{q,j}^{(0)}$, its appearance causes spectral weight loss from this original mode and new spectral weight confined to a limited momentum range. This implies that the polaron mode splits off from the bare lattice mode within a small finite momentum regime whereby the lattice mode seems to soften. Experimentally unusual mode softening has been observed for an almost dispersionless branch at high frequencies [25], where specifically the spectral weight transfer has been reported. Opposite to the interpretations offered in Ref. 25 in terms of phonon mode anomalies, we attribute the split off mode to the local polaron mode. Below T* the polaronic mode is stabilized, and remains dynamic with almost T independent energy (see Fig. 1). At $T_c$, however, a superconductivity-induced rapid mode softening sets in again if the energy gap has an s-wave component, whereas a pure d-wave gap leaves the polaron mode energy almost unchanged.

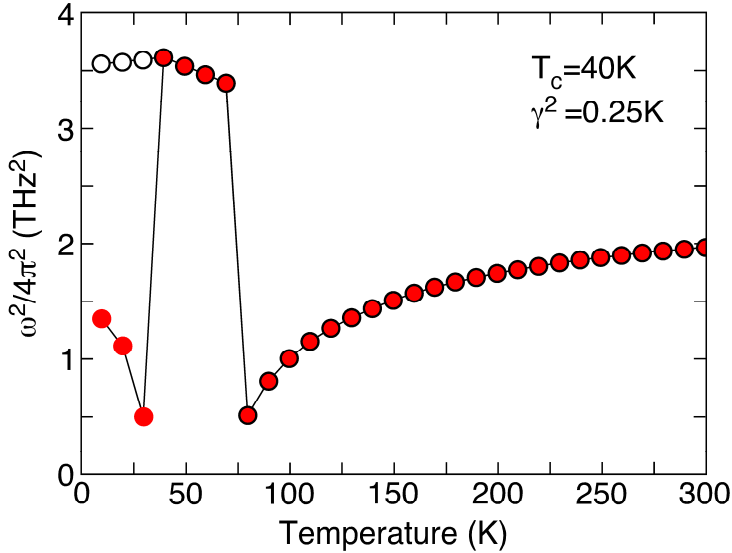

**Figure 1** Temperature dependence of the polaronic mode frequency for $\gamma^2 = 0.25$ and $T_c$=40K. The filled red circles / open black circles refer to an s-wave and a d-wave order parameter, respectively.

In order to compare these results with experimental data [4 – 6], we calculate the polaron mode related mean square lattice displacement via the fluctuation / dissipation theorem: $\sigma^2(T) = \hbar/(M\tilde{\omega}_{q,j})\coth(\hbar\tilde{\omega}_{q,j}/2kT)$, where $\tilde{\omega}_{q,j}$ is given by Eq. (4). In order to evaluate $\sigma^2(T)$ over the full temperature regime, i.e., for temperatures T>T*, $T_c$<T≤T* and T≤$T_c$, the opening of the superconducting gap also has to be taken into account. In analogy to the influence of the gap symmetry on the polaron mode frequency temperature dependence, here we again have the option of choosing the gap symmetry to be of d- or s-wave or the more complex s+d wave symmetry. In Fig. 2 we compare the mean squared displacements $\sigma^2(T)$ for the simpler cases of s- and d-wave order parameters only, and also show the effect of T* on $\sigma^2(T)$.

The first important feature is the sharp increase in $\sigma^2(T)$ at T* as a result of the local mode softening. Similar increases have been observed experimentally in EXAFS [4 – 6] and interpreted in terms of the appearance of two bond lengths in the $CuO_2$ planes with one corresponding to the undistorted lattice and the other to a distorted one [4]. At $T_c$, a strikingly different behaviour is observed for the two pairing symmetries investigated here: while for an s-wave gap another sharp peak appears at $T_c$, a d-wave gap does not have any significant influence on $\sigma^2(T)$. Comparing these findings with the experimental data [4 – 6], it must be concluded that the experiments closely trace an s-wave gap and are incompatible with a pure d-wave order parameter. However, coupled s+d order parameters as observed experimentally [22 – 24] exhibit the same qualitative features as the s-wave gap alone as long as the percentage contribution from the s-wave gap is larger than the one from the d-wave gap.

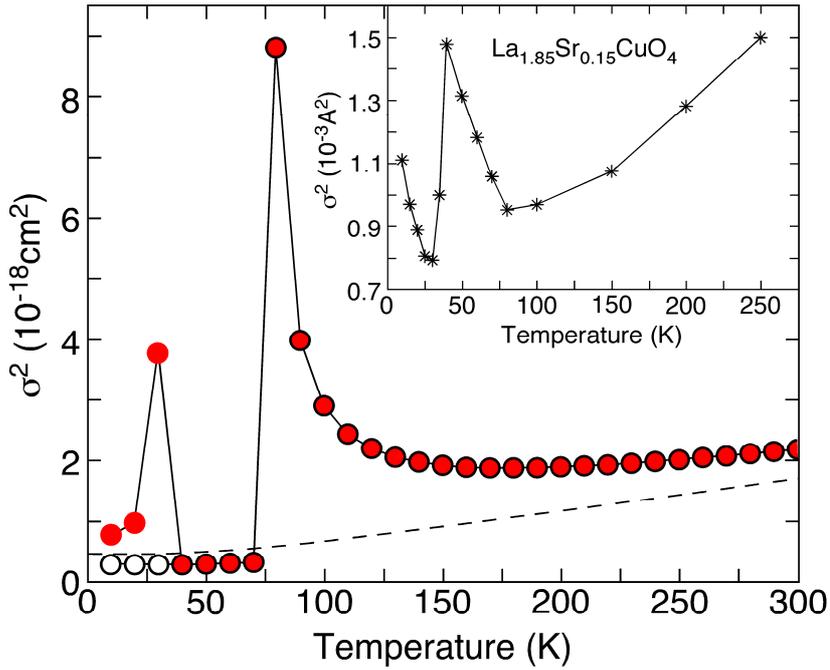

**Figure 2** The mean square displacement $\sigma^2(T)$ as a function of temperature for $\gamma^2 = 0.25$, and $T_c$=40K. The filled red circles / open black circles refer to an s-wave, respectively a d-wave order parameter. For comparison the bare unrenormalized mean square displacement ($\gamma = 0$) is added (dashed line). The inset shows the experimental data for $La_{1.85}Sr_{0.15}CuO_4$ of Ref. 26. Note, that in the calculations no damping or artificial line width broadening was introduced in order to minimize the number of parameters.

Our analysis of the mean-square displacement $\sigma^2(T)$ of the local distortions surrounding doped holes in terms of polaron formation captures the essential features observed by EXAFS experiments which is in agreement with results from quantum Monte Carlo simulations [27]. The coupled displacement of charge and lattice induces a local lattice mode renormalization with strong temperature dependent softening. This softening gives rise to a substantial increase in the mean square displacement in striking agreement with experiments. Below T*, the temperature at which the polarons form into a new dynamical pattern, superconductivity again affects $\sigma^2(T)$, but only if the order parameter is s-wave or mixed s+d wave: A pure or predominantly d-wave order parameter cannot reproduce the experimental findings. These data are thus compelling evidence, first, that *polarons survive in the superconducting phase*, and, second, that *an s-wave component of the order parameter must be present*, as already concluded from, e.g., muon spin rotation experiments [22 – 24]. Another important consequence of our calculation is the observation of a huge sign reversed isotope effect on T*, which we predict to be observable also by EXAFS experiments. One more important feature of the above analysis is that local probe experiments are able to differentiate between the symmetries of the superconducting order parameters if polaron formation is present.

**Acknowledgement** It is a pleasure to acknowledge stimulating discussions with H. Oyanagi, N. L. Saini and A. Bianconi. The work was supported by the K. A. Müller Foundation, the EU project CoMePhS and the NCCR program MaNEP.